\PassOptionsToPackage{numbers}{natbib}

\documentclass{article}

\pdfoutput=1

     \usepackage[preprint]{neurips_2019}

\usepackage[utf8]{inputenc} 
\usepackage[T1]{fontenc}    
\usepackage{hyperref}       
\usepackage{url}            
\usepackage{booktabs}       
\usepackage{amsfonts}       
\usepackage{nicefrac}       
\usepackage{microtype}      
\usepackage{caption}
\usepackage{mathtools}

\usepackage{physics}
\usepackage{bm}
\usepackage{subcaption}
\usepackage[algo2e]{algorithm2e} 
\usepackage{algorithm}
\usepackage{fancyhdr}
\usepackage{multirow}
\usepackage{xcolor}

\title{Beyond Observed Connections : Link Injection $^{*}$}

\makeatletter
\newcommand{\thickhline}{%
    \noalign {\ifnum 0=`}\fi \hrule height 0.65pt
    \futurelet \reserved@a \@xhline
}
\makeatother

\author{ Jie Bu \\
   Department of Computer Science\\
   Virginia Tech\\
   Blacksburg, VA 24061 \\
   \texttt{jayroxis@vt.edu} \\
  \And 
  M. Maruf \\
  Department of Computer Science\\
  Virginia Tech\\
  Blacksburg, VA 24061 \\
  \texttt{marufm@vt.edu} \\
  \And
  Arka Daw \\
  Department of Computer Science\\
  Virginia Tech\\
  Blacksburg, VA 24061 \\
  \texttt{darka@vt.edu} \\
  }
  
\begin{document}

\maketitle

\begin{abstract} 
\begin{quote}
In this paper, we proposed the \textit{link injection}, a novel method that helps any differentiable graph machine learning models to go beyond observed connections from the input data in an end-to-end learning fashion. It finds out (weak) connections in favor of the current task that is not present in the input data via a parametric link injection layer. We evaluate our method on both node classification and link prediction tasks using a series of state-of-the-art graph convolution networks. Results show that the link injection helps a variety of models to achieve better performances on both applications. Further empirical analysis shows a great potential of this method in efficiently exploiting unseen connections from the injected links.
 \end{quote}
\end{abstract}

\section{Introduction}
\label{sec:intro}
Recently, we have seen a growing popularity of graphs in machine learning\cite{kipf2016semi, hamilton2017inductive}. Even though deep learning can capture arbitrary patterns in the data, it often ignores the plethora of additional information which it can use if the data is represented as graphs. For example, in citation networks\cite{DBLP:journals/corr/YangCS16, sen2008collective}, the papers are linked to each other if one of the paper cites the other. This can be used to group papers in different categories. In protein-protein networks, a link between two proteins denote the type of interaction between them. These inherent structure in the data can be leveraged to learn richer node embedding, which in-turn allows us to improve predictive performance. Another drawback of traditional machine learning methods is that it assumes that the instances are independent of each other. Whereas for graphs, this assumption not valid, as the interaction between two nodes are incorporated in the edges, which might or might not be represented by a feature set.

Using graph to represent the knowledge we have, we can categorize graph data into three categories:

\begin{enumerate}
    \item The true graph: the actual graph that contains all the information for the target problem. We use $G^*$ or $G^*(X, A)$ to denote it, where $X$ is the node features and $A^*$ is the adjacency matrix.
    \item The observed graph: the partial graph that given by the data, can be viewed as the true graph with some edges being dropped-out. We use $G$ to denote it. In the link prediction problems, we reasonably assume the only missing part is part of the edges from $G^*$, therefore the node features stay the same, so we use $G(X, A)$.
    \item The predicted graph: the observed graph with predicted links. We use $\hat{G}$ or $\hat{G}(X, \hat{A})$ to denote it.
\end{enumerate}

Most existing methods try to learn structural patterns from the observed graphs in a supervised way then use these patterns to predict links, common state-of-arts include walk-based methods and network flow based methods. Recently graph neural networks (GNNs) and its convolutional variants - graph convolution networks (GCNs) attract lots of attentions in link prediction researches. These methods are more or less depend on some sort of message flows, e.g., for walk-based methods, the walks have starting points and transit to neighboring nodes at every step. The flow-based methods can be viewed as converged walks where randomness is addressed by statistical estimates. GNNs and GCNs are no exceptions as well. Both families rely on message passing to capture both input features and graph structures. To void confusions on the concepts and for convenience of describing the propagation of message flows in the networks, we refer such walks, flows and message propagation in the graph as message passing in this paper.

\begin{figure}[ht]
    \centering
    \includegraphics[width=0.58\textwidth]{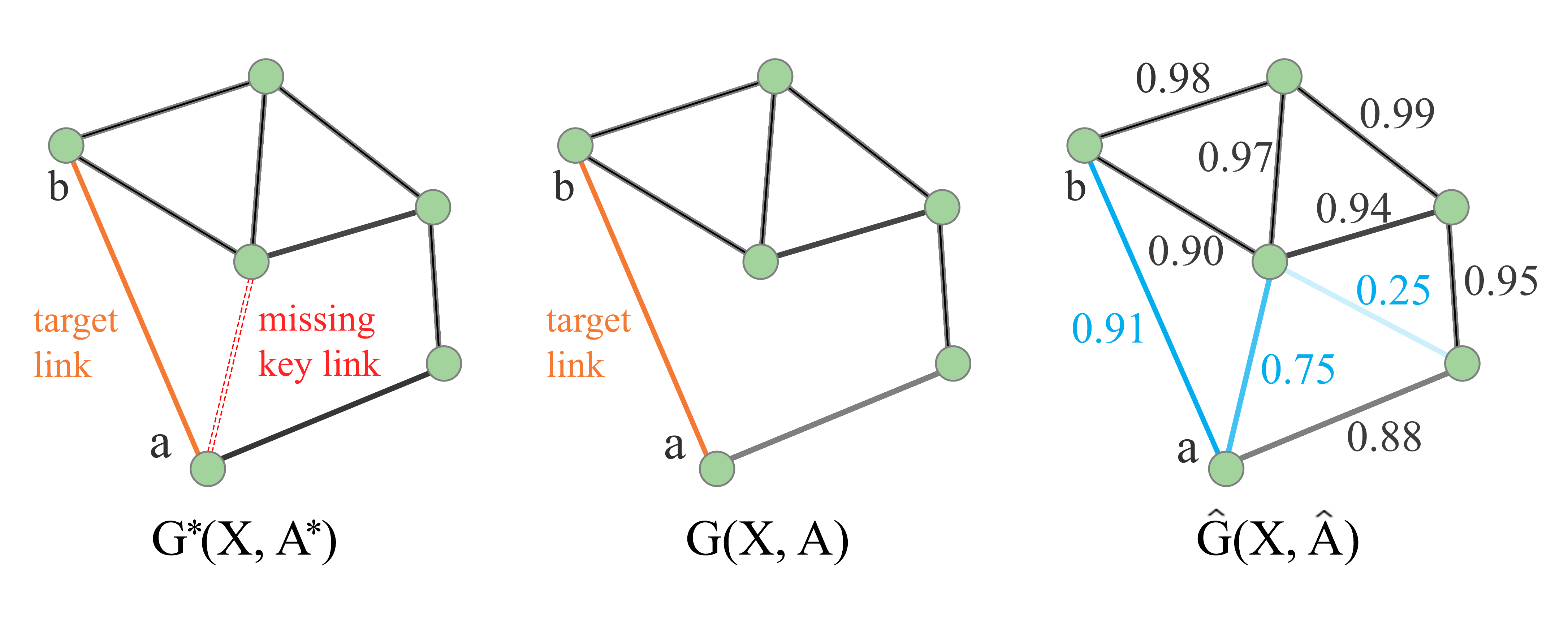}
    \caption{An Example of Missing Key Links (Red Dashed) And Injected Links (Blue Solid)}
    \label{fig:missing_key_link}
\end{figure}

We believe such message passing is crucial to estimate how strong the two nodes are connected. Previous researches on graphs empirically show that most real-world networks tend to have some clusters of nodes, i.e., connected nodes are more likely to belong to the same densely connected community or cluster of nodes. Based on the observed nature of real-world graphs, a reasonable model will tend to predict the presence of a link between two nodes when there exist a large amount of short paths and connections between them, in which case message can be passed more efficiently between the two nodes. This leads to a strong hypothesis that remains open for discussions and still needs future researches to prove: \textbf{for most of the state-of-the-arts, when message can not be effectively passed from a pair of nodes, the model will tend to predict the absence of a link between the two nodes}. The message passing mechanism integrated in most state-of-arts can effectively capture such connectivity information, allowing them to achieve good performances in real-world link prediction problems, e.g., social networks and protein and protein interaction (PPI).

However, one of the core idea of the paper that motivates of work is that the missing links destroy the neighboring connectivity that impair the message passing, hence lead to erroneously estimation about the connectivity strength between two nodes, especially when some of the key links are missed (see \textbf{Figure} \ref{fig:missing_key_link}).

The remainder of the paper is organized as follows. Section \ref{sec:background} provides insights into the related works in the broader domain of graph machine learning. Section \ref{sec:method} discusses our proposed method of link injection. Section \ref{sec:data} describes the datasets which we have used for evaluation. Section \ref{sec:exp_setup} provides additional information on the baselines and the evaluation metrics, while section \ref{sec:results} discusses in details our results. Section \ref{sec:conclusion} emphasises on our future work.
\section{Related Works}
\label{sec:background}
Our node classification and link prediction tasks is related to previous node embedding approaches, general supervised approaches to learning over graphs, and convolutional neural networks over graph-structured data.

\textbf{Embedding based approaches}: There are several node  embedding based approaches that learn low-dimensional embeddings using random walk statistics and matrix factorization based objectives. These embedding algorithms directly train node embeddings to individual nodes and also they require expensive additional training to make predictions over new nodes. Random walk based methods like DeepWalk \cite{perozzi2014deepwalk}, node2vec \cite{grover2016node2vec} uses the skipgram model used in word2vec \cite{mikolov2013distributed}, and represents the walks as sentences. These methods are related to classical approaches like spectral clustering \cite{ng2002spectral} as well as page-rank algorithms \cite{page1999pagerank}. Another type of embedding based approach comprises the Graph Auto-encoders (GAE) \cite{kipf2016variational}. GAEs are unsupervised learning frameworks- aim to learn low dimensional embedding via an encoder and then reconstruct the graph via a decoder. Graph autoencoder models tend to employ GCN 
as a building block for the encoder and reconstruct the structure information via a link prediction decoder. 

\textbf{Supervised learning over graphs}: GNN (Graph Neural Network) is a type of neural network which directly operates on the graph structure. Suppose for node classification task, each node is associated with a label, and we want to predict the label of the nodes without ground-truth. GNN leverages the labeled nodes to predict the labels of the unlabeled, and it does so using message passing or neighborhood aggregation. At each iteration it learns a low dimensional vector representation for each node that contains its neighborhood information. With the right choice of the loss function in the supervised setting, these GNNs can back propagate the loss and learn the task specific model. These methods \cite{dai2016discriminative, gori2005new, li2015gated, scarselli2008graph} have used GNN for node classification task. 

\textbf{Graph convolutional networks}: There are several convolutional neural network architectures for learning over graphs have been proposed in recent years \cite{bruna2013spectral, defferrard2016convolutional, duvenaud2015convolutional, kipf2016semi, niepert2016learning}. Some of the methods use spectral convolution. Defferrad et al \cite{defferrard2016convolutional} uses fast localized convolutions and kipf et al \cite{kipf2016semi} uses fast approximate convolutions in a semi-supervised setting. Both of the algorithms require to know the full graph Laplacian during training. GraphSAGE (SAmple and aggreGatE) \cite{hamilton2017inductive} an inductive node embedding approach can be viewed as an extension of these convolutional approaches. It aggregates the feature information from a node’s local neighborhood and by doing it simultaneously, it captures the topological structure of each node’s neighborhood. 

\textbf{Graph Connectivity Augmentation}: The idea of augmenting the graph connectivity for better message propagation in GCNs is not new, which shows it effectiveness in Graph U-Nets\cite{gao2019graph}. Similar to \cite{chepuri2016subsampling} which built links between nodes in a graph whose distances are at most $k$ hops, we also build "weak links" initially but for every pair of nodes. An rough analogy is that the methodology of link injection can be viewed as a learnable global graph connectivity augmentation. However, the idea of learning injected links is first in the community and substantially different to all existing graph connectivity augmentation. 

\section{Proposed Method : \textit{Link Injection}}
\label{sec:method}
To address the problems mentioned about, we propose \textit{link injection} as a way that retrieves part of the missing structural information in the given/observed graph. It provides a way of using artificial connections, i.e., injected links, to augment the message passing on the observed graph in the training phase. 

\begin{figure}[ht]
    \centering
    \includegraphics[width=1.00\textwidth]{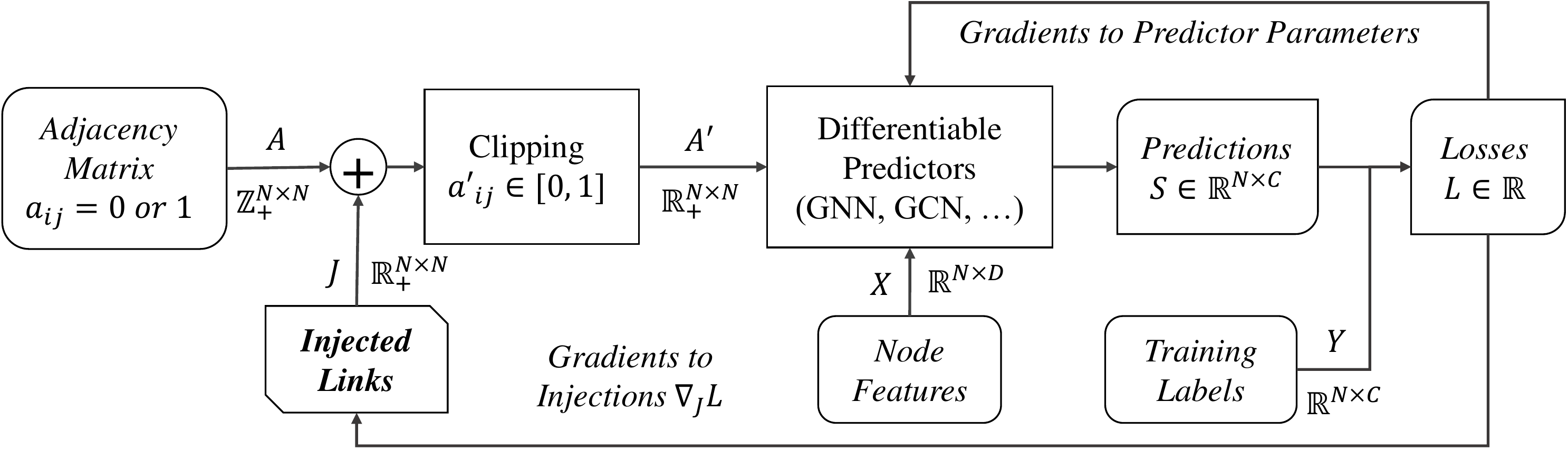}
    \caption{Flow Chart Showing The Pipeline of End-to-End Link Injection Architecture}
    \label{fig:flow_chart}
\end{figure}

Figure \ref{fig:flow_chart} shows the pipeline of the basic link injection method. A highlight of this method is that the connected strength of the injected links can be trained in an end-to-end fashion when combined with differential models, e.g., graph neural networks, etc. In practice, the injected links $J$ are often regarded 

A key for success using link injection is a proper constructed loss function since the connection strengths are updated in order to reduce the loss. This infers that there is no general way that can generate a certain set of injected links for any specific tasks, e.g., node classifications and link predictions, even on the same graph. When the loss functions can vary for different graph machine learning tasks, it means for different problems on the same graph, there may have different optimal sets of injected links that help our model to achieve best performances. Therefore, it is important to discuss the link injection methods under different contexts of graph machine learning problems. In the following sections, we will cover the discussions of link injections in two particular contexts, node classification and link prediction. 

\subsection{Link Injection For Node Classification}
First, we introduce the notations that will be used in the paper. Denote the loss function used for evaluating classification performances, e.g., cross-entropy loss functions, as $L$. Define a differential model $M: (\mathbb{R}^{N_i \times D}, \mathbb{R}^{N_i \times N_i}) \to \mathbb{R}^{N_i \times C}$, $i = 1, 2, ..., B$, where $B$ is the batch size or in many cases it will be the number of input graphs; $D$ is the dimension of the node features, $C$ is the number of output classes and $N_i$ is the number of nodes in the $i_{th}$ graph. The forward propagation of the model can be expressed as $M(X_i, A_i)$ given the input node features and adjacency matrix for the $i_{th}$ graph, and in the backpropagation phase, for the differential model $M$ the gradient calculated for model parameter $p$ is denoted as $\nabla_{p}{L}$. A example of a simple link injection for node classification tasks is shown in Algorithm \ref{algo:link_inj_clf}.

\begin{algorithm}[ht]
    \KwData{$G \to \{X, A, Y\}$; $X \in \mathbb{R}^{N \times D}$; $A \in \mathbb{R}^{N \times N}$; $Y \in \mathbb{Z}_{+}^{N \times 1}$, $\forall y \in Y, y \in [1, C]$}
    \KwResult{$S \in \mathbb{R}^{N \times C}$}
    \textbf{Parameter:} $J \in \mathbb{R}^{N \times N}$\; \\ 
    Initialize $J$ (random or constant)\; \\ 
    \While {Training} 
    {
        $A' \gets A + ReLU(J)$\\ \;
        $\text{Clip}(A')$ so that $A' \in [0, 1]$\\ \;
        $S \gets M(X, A')$\\ \;
        $L \gets \text{CrossEntropyLoss}(S, Y)$\\ \;
        $L \gets L + \sum_{p \in M}{\norm{p}} + \norm{J}$\\ \;
        $\nabla_{p}{L} = \text{Backpropagate}(M, L)$\\ \;
        $\nabla_{J}{L} = \text{Backpropagate}(J, L)$\\ \;
        $\text{Update}(M, \nabla_{p}{L})$\\ \;
        $\text{Update}(J, \nabla_{J}{L})$\\ \;
        \If{Termination Conditions Satisfied}{
            Stop training and return $M$. \;
        }
    }
    \caption{Link Injection For Node Classifications}
    \label{algo:link_inj_clf}
\end{algorithm}

By introducing additional connections, we hope link injection can help 

\subsection{Link Injection for Link Prediction}
\label{sec:linkpred}
Analogous to node classifications, the success for link predictions also depends highly on effective message passing, where link injection comes into help. It tries to maximize the likelihood of the observed graph for certain predicted links:
$$\max_{\hat{A}}{P(G|\hat{G})}$$
For the predicted links, we use a positive bias term to represent the difference between predicted graph and observed graph, $\hat{A} = A + J$. Then we can formulate the maximization problem as
$$\max_{J}{P(A|\hat{A}, X)}$$
where $J$ is a non-negative matrix.

To effectively obtain graph embedding for both node features and structural information, we choose GCNs as the predictor. Nevertheless, link injection works for a variety of state-of-art link prediction algorithms, so we use $F$ to denote the predictors, which map the input space to predicted links. Then we can use a simple formula to express the supervised model predicting as:
$$\hat{A} = F(A, X)$$
while our link injection method works in a semi-supervised fashion, the prediction gives a scoring matrix represents the estimated connection strength at each layer:
$$S = F(\hat{A}, X)$$
The loss function are designed as:
$$L = \norm{ReLU(A-S)}_F^2 + \lambda \norm{S}_F^2$$

\section{Datasets}
\label{sec:data}

We used the benchmark citation datasets - Cora and Citeseer \cite{DBLP:journals/corr/YangCS16, sen2008collective} for our experiments. In these network, nodes correspond to documents and edges correspond to citations. The feature of these networks are actually the bag-of-words of the documents. The subjects of the documents are represented as labels. The Cora dataset contains a number of Machine Learning papers divided into one of 7 classes while the CiteSeer dataset has 6 class labels. The stop words and the words that have frequency less than 10 have been removed. The final corpus of Cora has 2708 documents, 1433 distinct words in the vocabulary and 5429 links. And for CiteSeer, there are 3312 documents, 3703 distinct words in the vocabulary and 4732 links. In our preliminary experiment the link injector model outperforms some baselines in the node classification tasks.

Table \ref{table:1} shows the details of these datasets. 
\begin{table*}[t]
\centering
\scriptsize
\begin{tabular}{||c c c c c c||} 
 \hline
 DATASET & \#CLASSES & \#NODES & \#NODE FEATURES & \#EDGES & \#DEGREE(min, max, avg)\\ [0.5ex]
 \hline\hline
 Cora & 7 & 2,708 & 1,433 & 4,732 & (1, 168, 4) \\ 
 Citeseer & 6 & 3,327 & 3,707 & 5,429 & (1, 99, 3) \\
 \hline
\end{tabular}
\caption{Dataset Statistics}
\label{table:1}
\end{table*}

\section{Experimental Setup}
\label{sec:exp_setup}

\subsection{Node Classification experiments}
We evaluate our models on \textit{Cora} and \textit{CiteSeer} dataset for node classifications. For splitting the dataset, we used the common split given by \cite{DBLP:journals/corr/YangCS16}.
We use three different graph neural networks and graph convolutional networks as the differentiable predictors in Figure \ref{fig:flow_chart} to test the link injection method on different models. The three models are \textit{Graph Convolutional Networks} (GCNConv) \cite{kipf2016semi}, \textit{Graph SAmple and aggreGatE} (GraphSAGE) \cite{hamilton2017inductive}, and a simple graph neural network (GNN) under the message passing and neighboring aggregation framework as the other two models.

All models are trained for 10,000 epochs using Adamax optimizer \cite{kingma2014adam}. A sliding window early stopping trick was used to prevent overfitting. Basically it calculates the mean accuracy of the latest two non-overlapping windows on the validation set. If the mean accuaracy of the latter window is lower than the previous one by a margin that is larger than a set tolerance value, the training will stop. For our experiments, we set the sliding window size = 100, tolerance = 0.005, and earliest stop = 5000, i.e., no early stopping before the $5000_{th}$ epoch.
\subsection{Link Prediction experiments}
To evaluate our models for link prediction we have used the \textit{Cora} dataset. The baselines are same as those used for node classification. \footnote{All codes can be founded on \href{https://github.com/jayroxis/Link-Injection}{Github: \texttt{https://github.com/jayroxis/Link-Injection}.}}

\section{Results/Evaluation}
\label{sec:results}
\subsection{Comparing the effects of link injection for node classification}
Table \ref{tab:node_classification} compares the accuracy(macro) and area under the ROC curve(macro) for the different baselines on the Cora and CiteSeer datasets. Preliminary results of using link injection in Graph Neural Networks for node classification show significant improvement in the evaluation metrics for both datasets. For Cora, an improvement of 2.6\% in accuracy was observed while for CiteSeer dataset we observed an significant 6.7\% in accuracy improvement over the baseline GNN. Link injection in the GCN Conv improved accuracy for Cora and CiteSeer by 2.2\% and 2.\% respectively. Although the accuracy of GraphSAGE improved by 1.7\% on CiteSeer, we observed a slight decrease of 0.8\% in accuracy for Cora dataset. These results are coherent with our original hypothesis that link injection allows efficient message passing among graphs.      

Our preliminary experiments also show an all around improvement in Area Under the ROC curve. It can be inferred that link injection improves the degree of separability of the model, thus it is better at distinguishing between the classes than their respective baselines. This is also reflected in the improved accuracy for node classification as previously observed. 

\begin{table*}[ht!]
    \small
    \centering
    \begin{tabular}{c|c c|c c}
    \thickhline
    Dataset & \multicolumn{2}{c|}{Cora} & \multicolumn{2}{c}{CiteSeer} \\ \hline
    Metric & Accuracy (macro) & AUC-ROC (macro) & Accuracy (macro) & AUC-ROC (macro) \\ \hline
    GNN & $0.772\pm0.013$ & $0.881\pm0.004$ & $0.632\pm0.007$ & $0.765\pm0.004$ \\ 
    GNN* & \bm{$0.798\pm0.003$} & \bm{$0.895\pm0.002$} & \bm{$0.699\pm0.008$} & \bm{$0.809\pm0.003$} \\ \hline
    GCNConv & $0.757\pm0.003$ & $0.872\pm0.002$ & $0.660\pm0.010$ & $0.785\pm0.004$ \\ 
    GCNConv* & \bm{$0.772\pm0.008$} & \bm{$0.883\pm0.005$} & \bm{$0.687\pm0.011$} & \bm{$0.797\pm0.003$} \\ \hline
    GraphSAGE & \bm{$0.804\pm0.004$} & \bm{$0.893\pm0 .001$} & $0.680\pm0.004$ & $0.797\pm0.002$ \\ 
    GraphSAGE* & $0.796\pm0.001$ & $0.885\pm0.001$ & \bm{$0.697\pm0.001$} & \bm{$0.800\pm0.001$} \\ \thickhline
    \end{tabular}
    \caption{Results for node classification tasks on \textit{Cora} and \textit{CiteSeer} datasets.}
\label{tab:node_classification}
\end{table*}

In order to evaluate the quality of injected links and test our hypothesis, we designed an experiment where we train the models (both with and without link injection) on graphs where no information regarding the edges are available during training. As expected, due to unavailability of edges during the training on Cora, all the baseline models without link predicion performed very poorly. With 7 classes for Cora dataset, we anticipated an accuracy of $1/7 \approx 0.14$ which was reflected through our experiments as well. 

\begin{table*}[ht!]
    \small
    \centering
    \begin{tabular}{c|c c|c c|c c}
        \thickhline
         & GraphSAGE & GraphSAGE* & GCNConv & GCNConv* & GNN & GNN* \\ \hline
        Accuracy & 0.144 & \textbf{0.319} & 0.1300 & \textbf{0.316} & 0.1300 & \textbf{0.314} \\ 
        Hits & $\approx 0$ & \textbf{17} & $\approx 0$ & \textbf{15} & $\approx 0$ & \textbf{11} \\ 
        Hit Rate & $\approx 0$ & \textbf{0.161\%} & $\approx 0$ & \textbf{0.142\%} & $\approx 0$ & \textbf{0.104\%} \\ 
        MR & N/A & \textbf{4214} & N/A & \textbf{5737} & N/A & \textbf{5243} \\ 
        \multicolumn{1}{l|}{MR Ratio} & $\approx 0$ & \textbf{0.601} & $\approx 0$ & \textbf{0.457} & $\approx 0$ & \textbf{0.503} \\ \thickhline
    \end{tabular}
    \caption{Evaluating injected links in node classification task on \textit{Cora} with \textbf{no edges available in training}. Accuracy is the macro-accuracy for node classification while Hits Hit Rate and MR are metrics for evaluating the injected links against observed connections in the graph. The best results among five random experiments will are reported in the table.}
\label{tab:no_edges}
\end{table*}

\subsection{Comparing the effects of link injection for link prediction}

Table \ref{tab:link_prediction} compares the accuracy, precision and recall for link prediction for the different baselines on the Cora dataset. The results show significant improvement in accuracy and precision for baselines with link injection. The recall also improves for GNN, while we observe a slight decrease in recall for GCNConv and GraphSAGE when equipped with link injection. Overall, the results show by introducing weak artificial links during the training process, and learning the weights of the weak links, the overall performance of the baselines for link prediction tasks can be improved significantly.
\begin{table*}[ht]
\small
\centering
\begin{tabular}{c|c c c}
\thickhline
Model & Accuracy (\%) & Precision (\%) & Recall (\%) \\ \hline
GNN & $88.71\pm0.59$ & $95.92\pm0.92$ & $80.87\pm0.54$ \\
GNN* & \bm{$92.56\pm1.69$} & \bm{$97.46\pm0.87$} & \bm{$87.42\pm3.90$} \\ \hline
GCNConv & $93.48\pm0.29$ & $95.53\pm0.52$ & \bm{$91.25\pm0.70$} \\
GCNConv* & \bm{$93.99\pm0.26$} & \bm{$98.34\pm0.62$} & $89.49\pm0.15$ \\ \hline
GraphSAGE & $92.33\pm0.37$ & $91.72\pm0.66$ & \bm{$93.07\pm0.13$} \\
GraphSAGE* & \bm{$93.29\pm0.45$} & \bm{$96.32\pm1.10$} & $90.02\pm0.22$ \\ \thickhline
\end{tabular}
\caption{Results for link prediction tasks on \textit{Cora} datasets.}
\label{tab:link_prediction}
\end{table*}

Table \ref{tab:injected_link_neighborhood} demonstrates the injected links in the neighborhood from link prediction tasks on Cora dataset. To demonstrate that injected links learned by our model are not arbitrary, we perform some analysis on the learned injections. When the top 50/10556 links were used for evaluating the injected links, it was found that for all the baselines, the injected links were either neighbors(i.e. an edge was present between them in the original graph) or they were completely disconnected(i.e. there were no path between the two nodes). 

%

\begin{table*}[ht]
\centering
\small
\begin{tabular}{l|rr|rr}
\thickhline
\multicolumn{1}{c|}{\multirow{2}{*}{Models}} & \multicolumn{2}{c|}{Top 50} & \multicolumn{2}{c}{Top 10556} \\
\multicolumn{1}{c|}{} & \multicolumn{1}{r}{Neighbors}  & Disconnected & \multicolumn{1}{r}{Neighbors} & Disconnected \\ \hline
GCNConv* & 100.00\% & 0.00\% & 60.72\%  & 39.28\% \\ 
GNN* & 100.00\%  & 0.00\% & 86.92\%  & 13.08\% \\ 
GraphSAGE* & 10.00\% & 90.00\% & 72.95\% & 27.05\% \\ \hline
\end{tabular}
\caption{Injected links in its neighborhood from link prediction task on the \textit{Cora} dataset.}
\label{tab:injected_link_neighborhood}
\end{table*}

Since, Cora dataset has 10556 edges, we used the top 10556 values of the injected links, and evaluated them with the edges already present in the Cora dataset to assess the quality of the injected links. Table \ref{tab:injected_link_stat} show the statistics of 10556 top-scored injections.


\begin{table*}[ht!]
\centering
\small
\begin{tabular}{l|lll}
\thickhline
Models & GCNConv* & GraphSAGE* & GNN* \\ \hline
Training Fraction & 80\% & 80\% & 80\% \\ 
Hits (Total) & 6410 & 7701 & 9175 \\ 
Hits ($\not \in$ Train) & 182 & 101 & 403 \\ 
Hit Rate (Total) & 60.724\% & 72.954\% & 86.917\% \\ 
Hit Rate ($\not \in$ Train) & 8.621\% & 4.784\% & 19.089\% \\ 
MR & 2567 & 3829 & 2341 \\ 
MR Ratio & 0.7568 & 0.6372 & 0.7782 \\ \thickhline
\end{tabular}
\caption{Statistics of 10556 top-scored injected links in reflecting the original (observed) graph.}
\label{tab:injected_link_stat}
\end{table*}
\section{Discussion}
\label{sec:discussion}

\begin{figure}[ht]
\centering
\begin{subfigure}[b]{.30\linewidth}
\includegraphics[width=\linewidth]{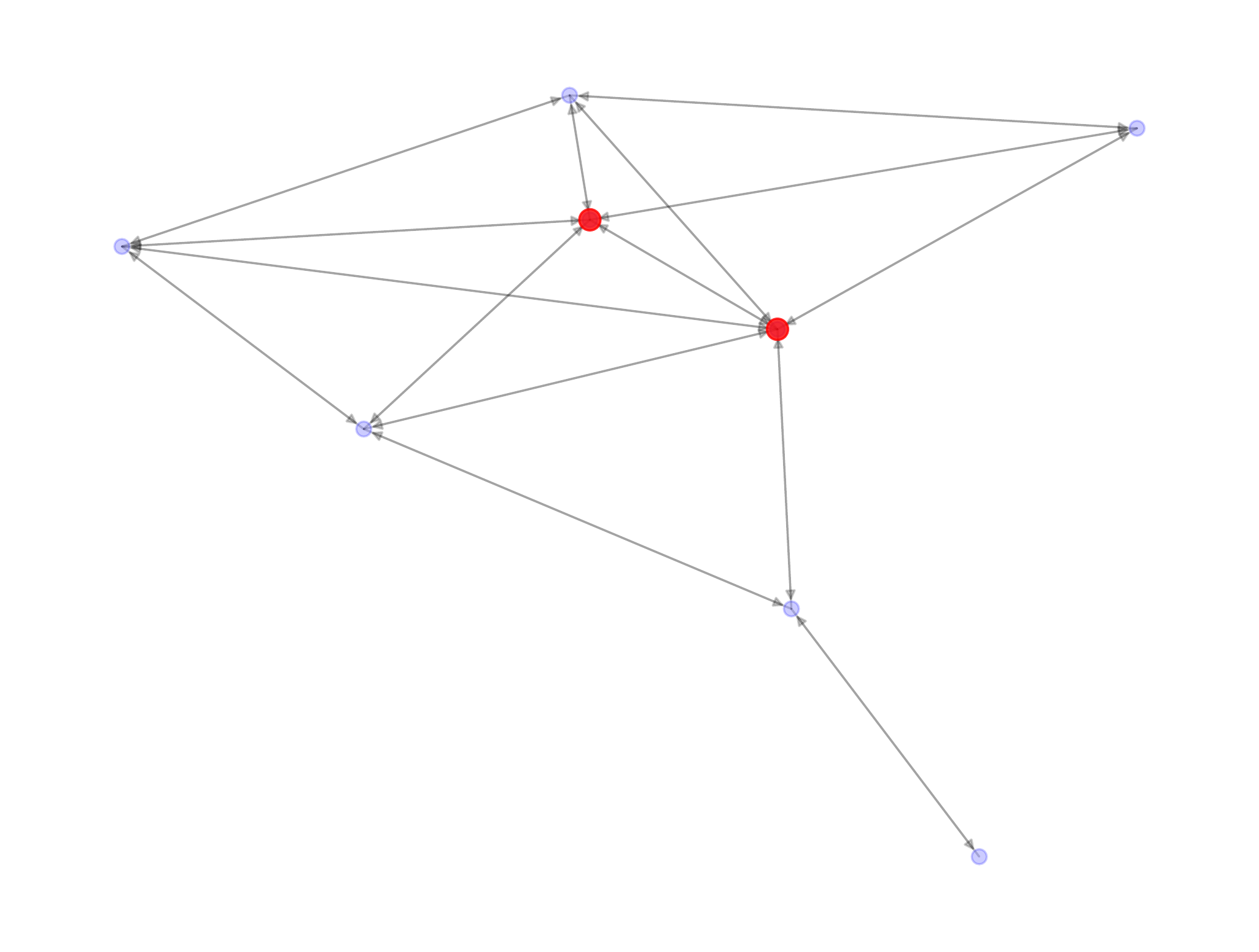}
\caption{GCNConv}\label{fig:GCN_top_injected}
\end{subfigure}
\begin{subfigure}[b]{.30\linewidth}
\includegraphics[width=\linewidth]{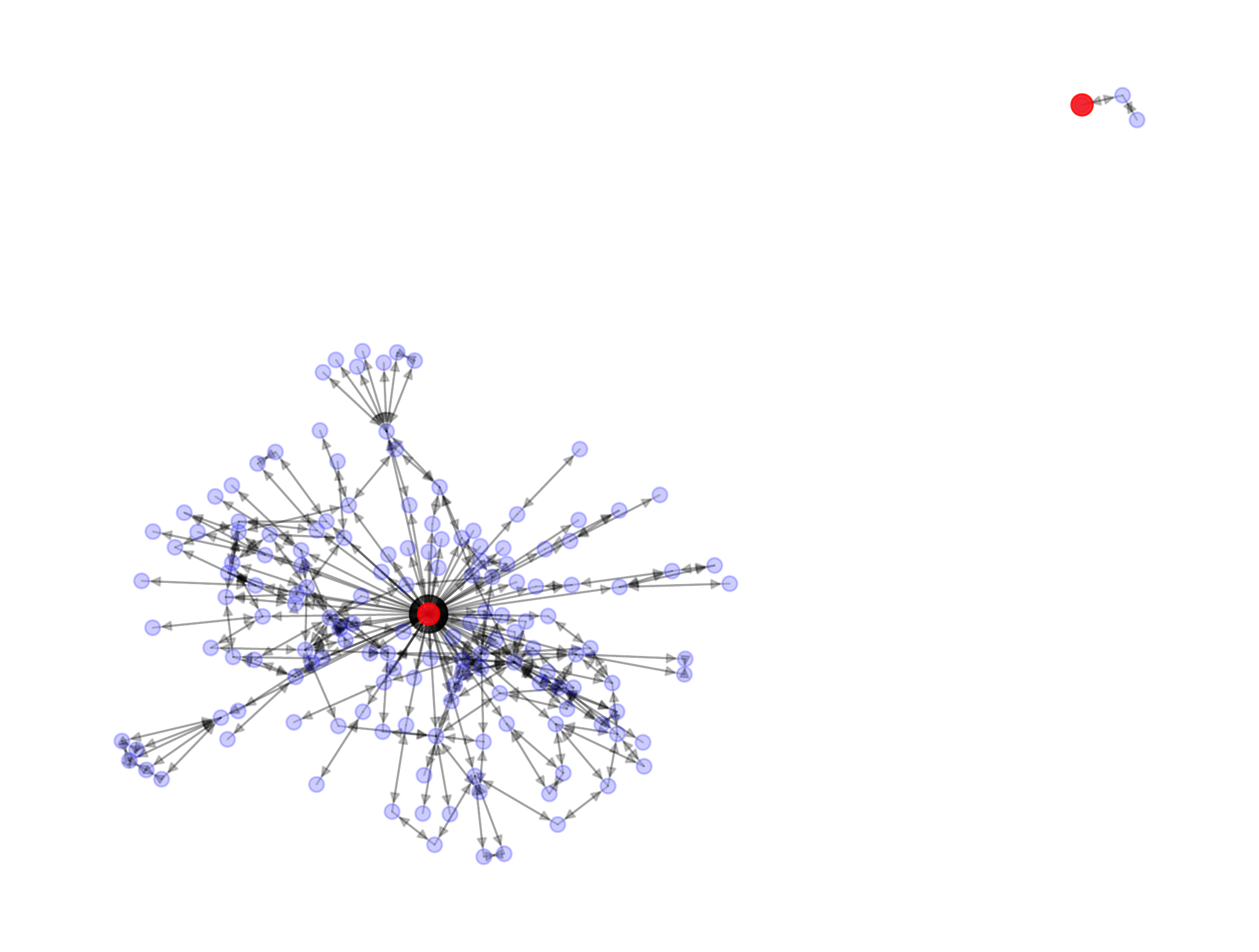}
\caption{GraphSAGE}\label{fig:SAGE_top_injected}
\end{subfigure}
\begin{subfigure}[b]{.30\linewidth}
\includegraphics[width=\linewidth]{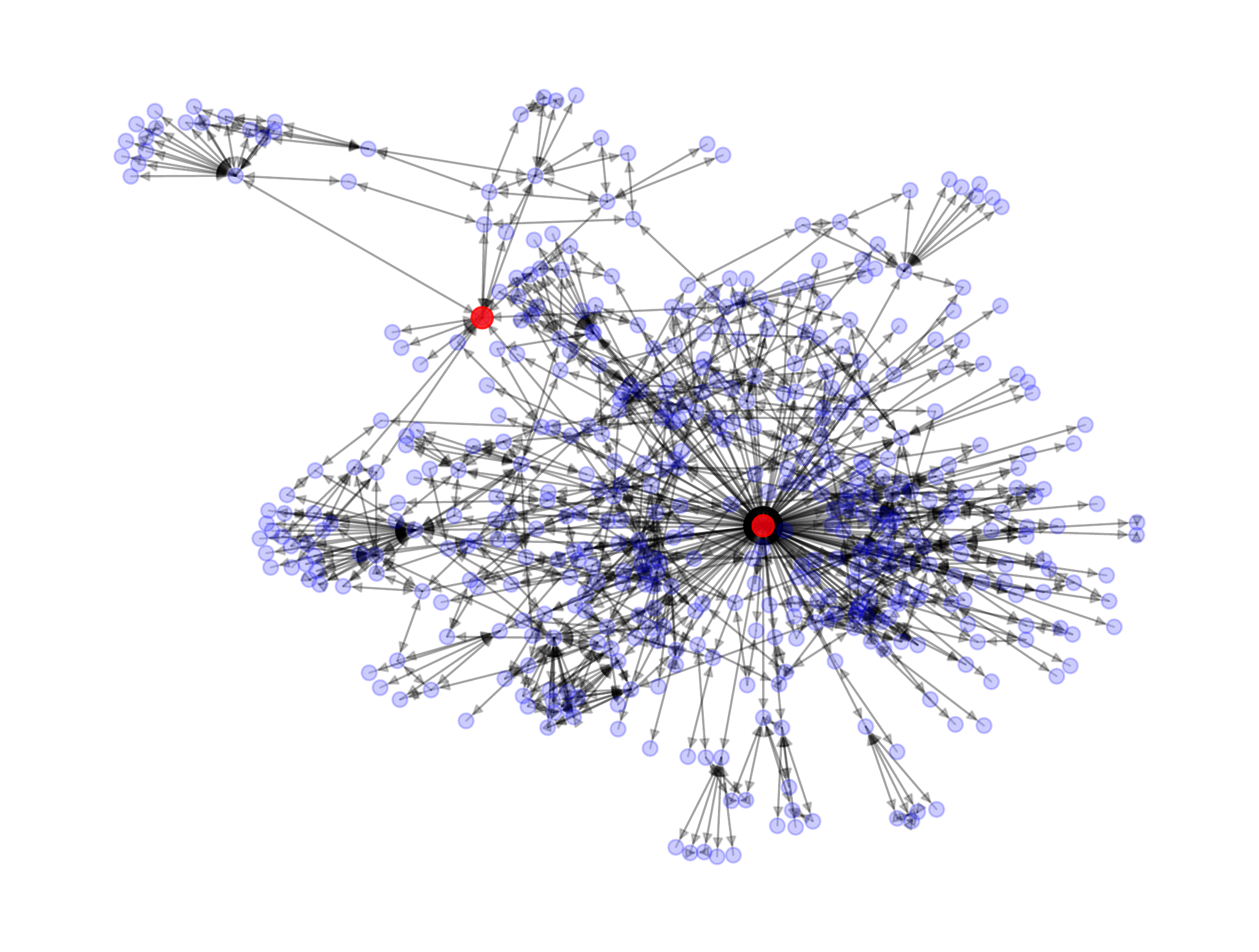}
\caption{GNN}\label{fig:GNN_top_injected}
\end{subfigure}
\caption{Examples of the top-scored injected links of different models in the link prediction task on the \textit{Cora} dataset.}
\label{fig:top_1_injection_link_pred}
\end{figure}

Apart from the results showing in section \ref{sec:results}, we found more interesting observations. Here we include several things to discuss about.
\begin{enumerate}
    \item The link injection displayed incredible consistency with the original graphs, especially in the link prediction task.
    \item Different predictive models can result in different preferences in injections.
\end{enumerate}

\begin{figure}[ht!]
\centering
\begin{subfigure}[b]{.485\linewidth}
\includegraphics[width=\linewidth]{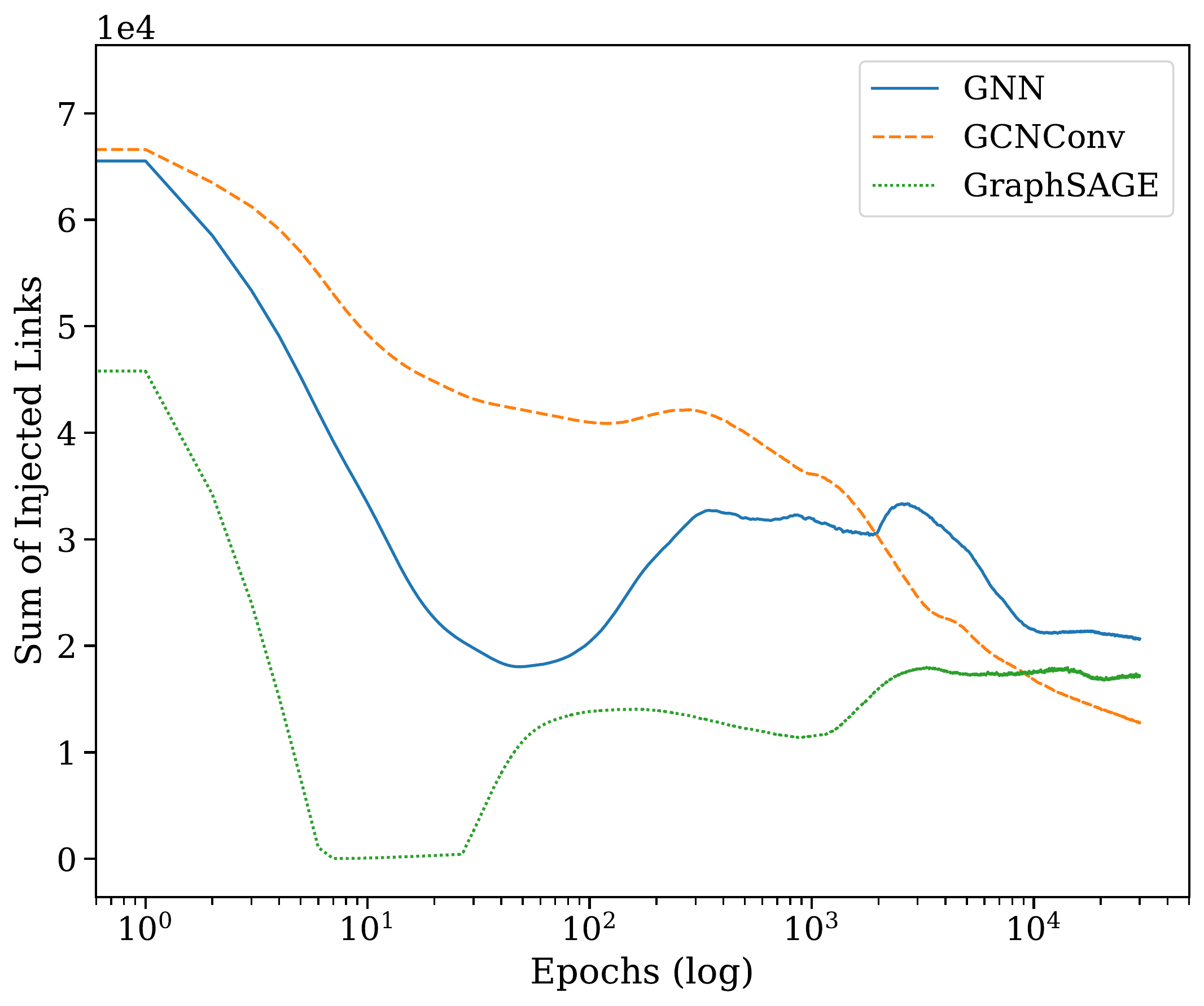}
\caption{Sum of Injections Per Epoch}
\label{fig:sum_injection_per_epoch}
\end{subfigure}
\begin{subfigure}[b]{.50\linewidth}
\includegraphics[width=\linewidth]{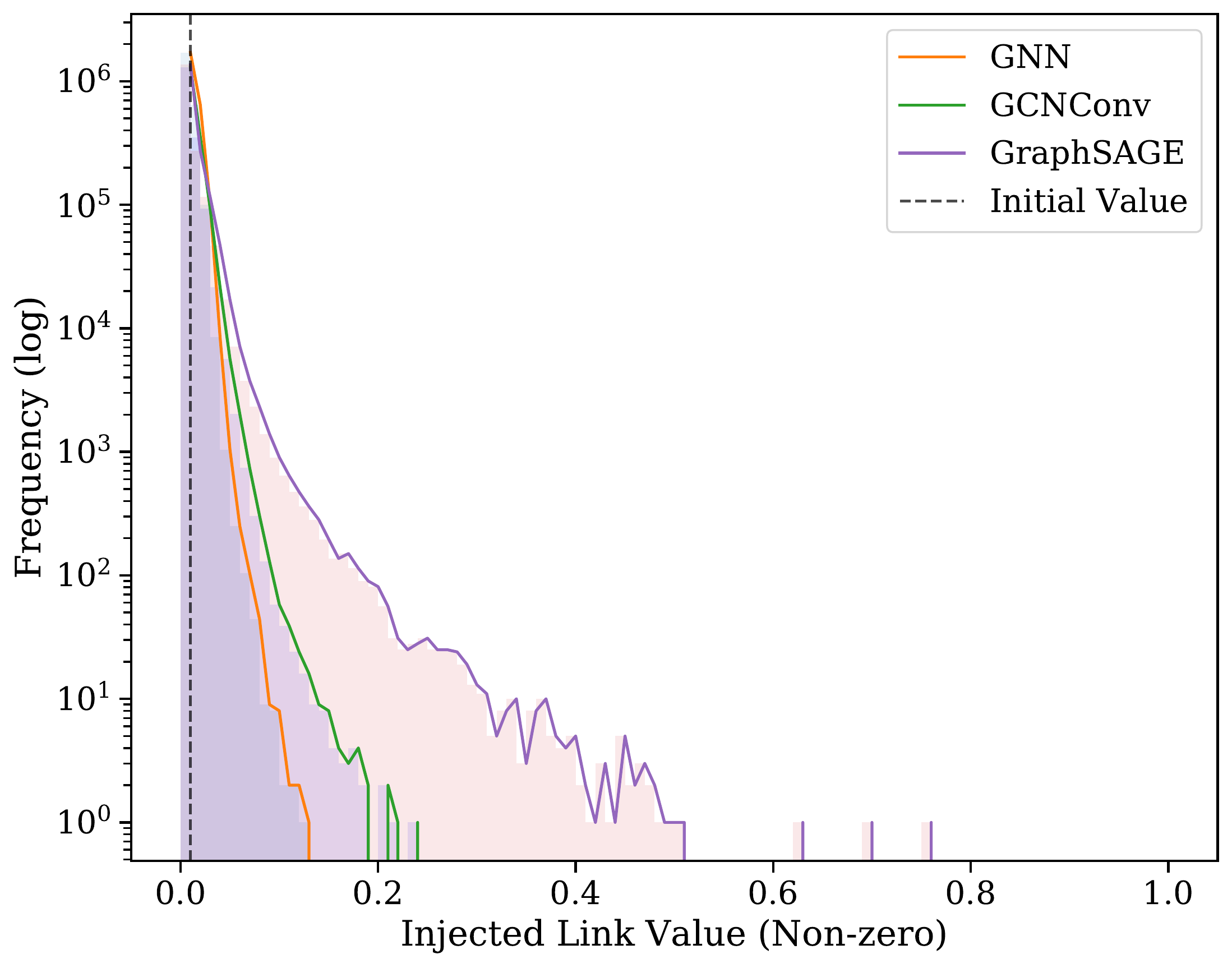}
\caption{Distribution Plot of Non-zero Injections}
\label{fig:dist_injection_nonzero}
\end{subfigure}
\caption{Examples of the top-scored injected links of different models in the link prediction task on the \textit{Cora} dataset.}
\label{fig:stat_injection}
\end{figure}

Figure \ref{fig:top_1_injection_link_pred} shows the neighboring subgraph of the highest-scored injected link of different models in the link prediction task on \textit{Cora}, where the red dots are the vertices of the injected link and blue dots are the nodes in the neighboring subgraph. The black solid lines are the edges that connect these nodes. A noticeable difference between the GraphSAGE and GCNConv or GNN is that the GraphSAGE tended to welcome the assistance of the injected links that lay between disconnected communities, while the  It probably acts as a factor causing GraphSAGE we constructed with link injection hard to be trained.


Figure \ref{fig:sum_injection_per_epoch} shows that the injections learned during the training process is different for different baselines, but the sum of the injected links for the different baselines seems to converge after the training is complete. Also, figure \ref{fig:dist_injection_nonzero} shows the distribution of the injections learned by the model. It can be seen that the value of the injections and the number of injections are inversely proportional. Furthermore, it must be noted that a large number of injections become zero after the training process is finished, which is in accordance to our goals, i.e. we want the injections to be sparse. 
\section{Future Work}
\label{sec:conclusion}
Link injection when paired with differentiable predictors such GNN, GCNConv, and GraphSAGE show significant improvement in performance for node classification and link prediction tasks. Also, the learned injections are not arbitrary and demonstrate interesting properties as described earlier. As an extension to this work, we plan on implementing negative links injections to deal with noisy links which might be present in the data. Since, a major drawback of our proposed approach is the computational overhead of using the adjacency matrix, a possible extension of this work would be using diff-pool \cite{ying2018hierarchical} to reduce the size of the graph and construct link injections on the coarsened graph. Other possible extensions include storing a different injection matrix for each graph when the problem involves multiple graphs.

\bibliographystyle{unsrt}
\bibliography{reference}

\end{document}